# Ion Chambers for Monitoring the NuMI Neutrino Beam at Fermilab

Dharmaraj Indurthy,[1] Ryan Keisler,[1] Sacha Kopp,[1†] Steven Mendoza,[1] Marek Proga,[1] Zarko Pavlovich,[1] Robert Zwaska,[1] Deborah Harris,[2] Alberto Marchionni,[2] Jorge Morfin,[2] Albert Erwin,[3] Huicana Ping,[3] Christos Velissaris,[3] Donna Naples,[4] Dave Northacker,[4] Jeff McDonald,[4] Milind Diwan,[5] Brett Viren[5]

[1]*Department of Physics, University of Texas, Austin, Texas 78712*
[2]*Fermi National Accelerator Laboratory, Batavia, IL 60510*
[3]*Department of Physics, University of Wisconsin, Madison, Wisconsin 53706*
[4]*Department of Physics, University of Pittsburgh, Pittsburgh, Pennsylvania 15260*
[5]*Brookhaven National Laboratory, Upton, Long Island, New York 11973*

**Abstract.** The Neutrinos at the Main Injector (NuMI) beamline will deliver an intense muon neutrino beam by focusing a beam of mesons into a long evacuated decay volume. The beam must be steered with 1 mRad angular accuracy toward the Soudan Underground Laboratory in northern Minnesota. We have built 4 arrays of ionization chambers to monitor the neutrino beam direction and quality. The arrays are located at 4 stations downstream of the decay volume, and measure the remnant hadron beam and tertiary muons produced along with neutrinos in meson decays. We review how the monitors will be used to make beam quality measurements, and as well we review chamber construction details, radiation damage testing, calibration, and test beam results.

## INTRODUCTION

The NuMI beamline [1,2] at Fermilab will deliver an intense $\nu_\mu$ beam to the MINOS detectors at FNAL and at the Soudan Laboratory in Minnesota. Additional experiments are foreseen. The primary beam is fast-extracted onto the NuMI pion production target in 8.56 μsec spills from the 120 GeV FNAL Main Injector. The beam line is designed to accept $4\times10^{13}$ protons/spill. After the graphite target, two toroidal magnets called "horns" [3] sign-select and focus forward the secondary mesons (pions and kaons) into a 675 m volume evacuated to ~1 Torr to reduce pion absorption, where they may decay to muons and neutrinos. The horns and target may be positioned so as to produce a variety of neutrino beam energies [4].

The instrumentation foreseen to ensure the neutrino beam quality on a spill-by-spill basis combines primary, secondary, and tertiary beam measurements. The primary beam direction is monitored by segmented foil secondary emission monitors [5] and by capacitive beam position monitors. Its intensity is measured by a beam current toroid (BCT). Additionally, the NuMI target is electrically isolated, so may be read out as a Budal-type monitor [6] of the primary beam's fraction on target. The above instrumentation measures the quality of the beam reaching the NuMI target.

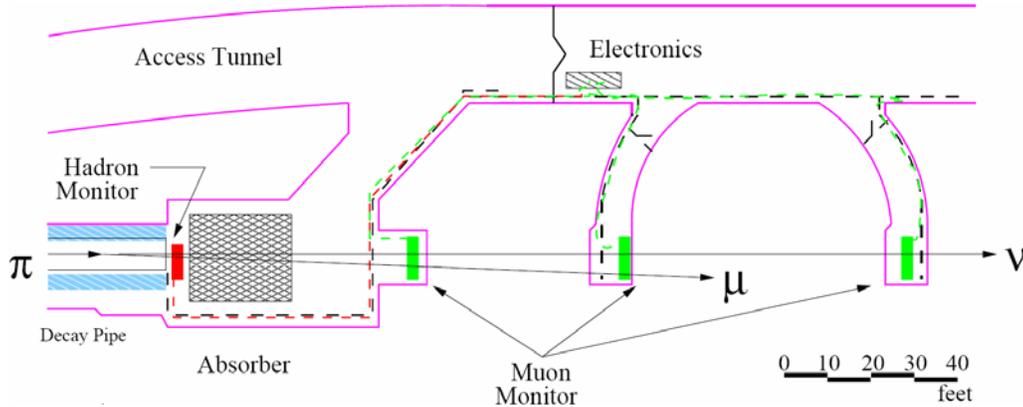

**FIGURE 1.** Layout of the secondary and tertiary beam monitors. Downstream of the decay volume, the hadron monitor measures flux and spatial profiles of remnant hadrons. At three stations in the downstream rock, the muon monitors measure rates and spatial profiles of the muon beam. The NuMI target and horns are 675m upstream, to the left of this figure.

The subject of this paper is the secondary and tertiary beam monitoring system, shown in Figure 1. Its purpose is to monitor the integrity of the NuMI target and of the horns which focus the secondary meson beam. This monitoring is accomplished by measuring the remnant hadron beam reaching the end of the decay tunnel and by measuring the intensity and lateral profile of the tertiary muon beam penetrating the beam absorber and the downstream dolomite rock. Because every muon is produced by the pion decays which produce neutrinos, the muon beam provides a good measure of the focusing quality of the neutrino beam.

## SECONDARY + TERTIARY BEAM MEAUREMENTS

The hadron monitor sits downstream of the NuMI decay volume, immediately in front of the beam absorber. It measures approximately 1m×1m, and is segmented to measure the spatial profile of particles arriving at the detector. Monte Carlo calculations indicate that the charged particle rate at the center of the decay pipe is dominated by protons which did not interact in the target.

Figure 2 shows the proton beam profile at the end of the decay volume. A prominent peak is visible at beam center, due to protons which did not react in the 1m graphite NuMI target. The 20cm width of this peak is expected, given the mean scattering angle of 120 GeV protons passing through the target and the 725 m drift distance from the target to the hadron monitor. During normal running, this scattering width serves as a monitor that the primary beam is centered on the NuMI target. As noted in Figure 2, it also monitors the integrity of the segmented NuMI target, since just 1 fin breaking off the target results in a more peaked proton spot. During initial beam commissioning, the hadron monitor helps align the primary proton beam to within 3cm/725m=42 µRad angular precision. This compares well to the 14 µRad provided by the pre-target SEM's, which only periodically are inserted in the beam.

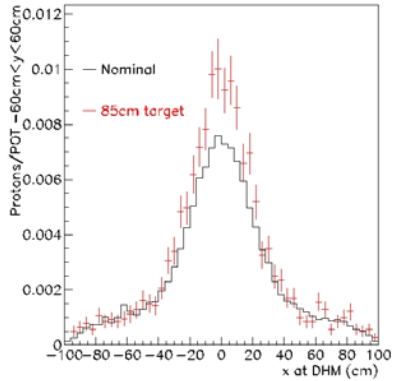

**FIGURE 2.** Horizontal profile of protons arriving at the downstream hadron monitor, over the band in the vertical direction of ±60cm about beam center. The profile is shown for two cases, normal beam operations with a 1m target, and the case in which the first target fin has broken off. The error bars are due to Monte Carlo statistics.

The 3 muon monitors are located downstream of the beam absorber. Because of the steel and rock upstream of each alcove, they detect muons of increasingly higher momentum, hence detect only the decay products of higher energy parent pions. In the lowest energy ν beam, the parent pions have energies of 4-5 GeV, so many of the tertiary muons do not penetrate into Alcove 0. In the higher energy beams, the horn-focused pions have energies 12-30 GeV, so the muons penetrate into Alcoves 0, 1 and 2. Multiple scattering through the rock broadens the beam as it traverses further downstream into Alcoves 1 and 2.

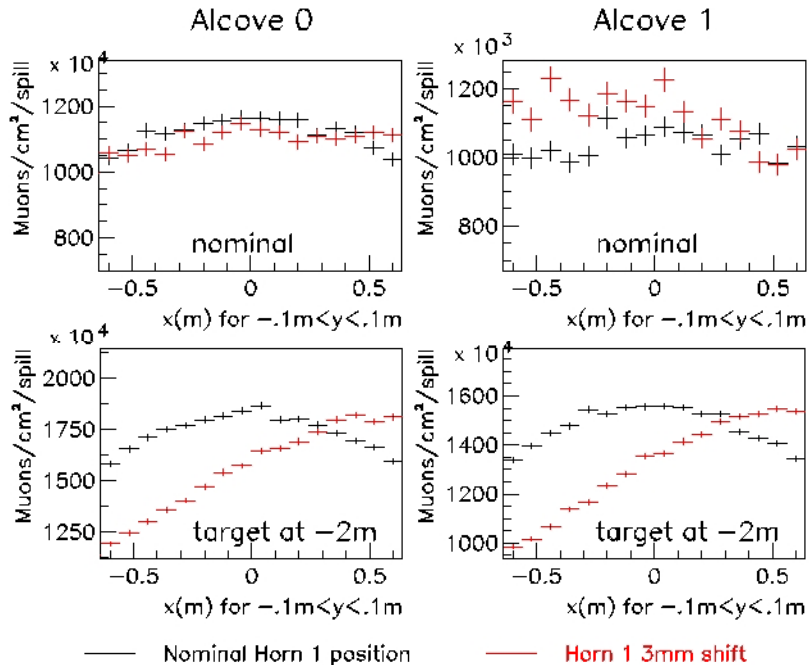

**FIGURE 3.** Monte Carlo calculations of muon beam profiles in Alcove 0 and Alcove 1. The upper plots are for low-energy beam, the two lower plots are for the highest neutrino beam energy. Shown are the profiles for normal beam (black points) and for the case in which the beam is mis-steered due to an offset of horn #1 by 3mm (red points). The error bars are due to limited Monte Carlo statistics.

**TABLE 1. Properties of the four ion chamber stations**

| Element | Muon Monitors (Low Energy Beam) | | | Muon Monitors (High Energy Beam) | | | Hadron Monitor |
|---|---|---|---|---|---|---|---|
| | Alcove 0 | Alcove 1 | Alcove 2 | Alcove 0 | Alcove 1 | Alcove 2 | |
| Distance from target (m) | 739 | 751 | 769 | 739 | 751 | 769 | 725 |
| Charged Particle Flux ($10^7/cm^2$/spill) | 1.1 | 0.23 | 0.09 | 3.0 | 1.7 | 0.22 | 2500 |
| Radiation Level ($10^6$ Rad/yr.) | 6.6 | 0.7 | 0.15 | 18 | 10 | 0.7 | ~2000 |
| Parent Particle Threshold (GeV) | 4 | 10 | 18 | 4 | 10 | 18 | <1 |
| μ Beam Flux(r=0)/ Flux(r=1m) | 1.3 | 1.1 | 1.1 | 2.1 | 1.5 | 1.5 | -- |
| μ Beam Shift for Horn 1 Offset (cm) | 2.5 ±1.5 | 0.6 ±1.5 | 1.3 ±1.5 | 23.0 ±1.5 | 12.8 ±1.5 | 3.3 ±1.5 | -- |

Table 1 summarizes some parameters of the muon alcoves. Listed are the distance from the target, particle flux, and radiation levels. Furthermore, Table 1 indicates the minimum pion energy whose daughter muon can arrive at an alcove; for the low-energy (high-energy) ν beam, the desired pion beam energy is 4 (40) GeV. As a demonstration of the fact that the muon beam becomes more sharply peaked at high ν energy, Table 1 lists the flux on beam axis ($r=0$) divided by the muon flux at the edge of the monitors ($r=1m$). Lastly, the table indicates the amount by which the muon beam centroid will shift in the scenario where horn 1 is misaligned by 3 mm. Figure 3 shows the muon beam profiles for a normal beam as well as a misaligned horn beam.

Because of the broad muon profile in the low energy beam, the muon monitors can monitor only relative neutrino event rate spill-to-spill when running in that configuration. The beam centroid shifts expected are barely measurable. Periodic short runs are therefore envisioned using the higher energy neutrino beams, which can be accomplished by moving the NuMI target remotely. Such a re-configuration thus allows a measure of the alignment of the horn optics on a periodic basis.

## ION CHAMBER CONSTRUCTION

Because of the large anticipated particle fluences, signal strength will not be a significant issue. In fact, the greater consideration will be to reduce space-charge build-up in the gas volume: slow-moving positive ions will develop their own electric field which screens the applied electric field of the chamber. Such screening increases the drift time across the gas volume, increasing the probability for charge recombination in the gas. Such loss of charge manifests itself as a non-linearity in ion chamber response *vs* the incident particle flux. This consideration motivates the use of He gas, which produces the lowest ionization density of any of the inert gases [7,8], thereby reducing the space charge effect

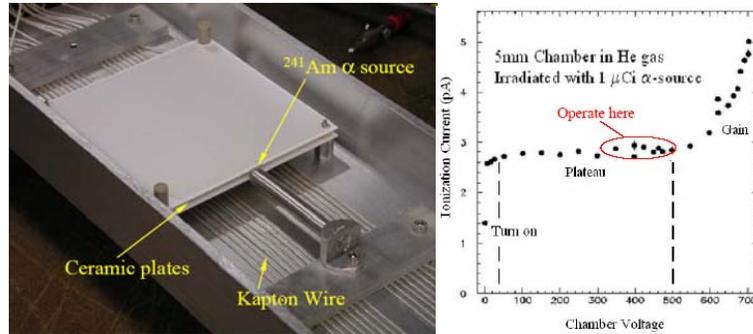

**FIGURE 4.** Ceramic parallel plate ion chambers. (left) Each chamber has an alpha source for calibrations. Kapton cables bring HV in and signals out. (right) Bias curve for a single chamber illuminated by the alpha source.

The ion chambers are parallel plates made from ceramic wafers with Ag-Pt electrodes [9]. The electrode separation is 3 mm (0.8 mm) for the muon (hadron) monitors. Because of the need to calibrate and track the relative responses of the 243 chambers in the muon arrays, each chamber is illuminated by a 1 µCi $^{241}$Am α source ($E_\alpha$=5.5 MeV), as shown in Figure 4. The ionization current from such a source in pure He gas is typically 2-4 pA, which may be measured between spills to monitor gas quality, pressure, or temperature variations (see Figure 4).

The hadron monitor is an array of 49 chambers mounted in a single Aluminum vessel. Each chamber requires 2 ceramic feedthroughs [10], one for signal readout and one for a bias voltage (see Figure 5). The vessel is sealed with an Indium wire, whose melting point of 150°C is well above the 50-60°C temperature at which the monitor will operate. The design maximizes use of Aluminum over stainless steel in order to reduce the presence of long-lived radionuclides in the detector [11]: with its current proportion of 54lbs Aluminum and 4lbs. Stainless, residual activation will be 58Rem/hr. The signal and HV lines are custom-made "cable" constructed of aluminum welding rod core, insulated by a ceramic tube, shielded by an aluminum sheath. After 1 m this cable transitions to a kapton-insulated cable [10].

The muon monitor stations are each 81 chamber arrays. The layout of 9×9 ion chambers is achieved by mounting 9 "tubes" onto a support structure vertically. Each tube contains a tray of 9 ion chambers, as shown in Figure 6. Because of the lower anticipated radiation levels in the muon alcoves, we routed the signals out to one end of the tube, where all feedthroughs come out one endplate of the tube. The tube is made from 6"×2" rectangular extrusion, into which a "U"-channel 'tray' is inserted. The endplates seal to the tubes using a soft Aluminum wire gasket. The HV feedthroughs are custom-made using PEEK tubes and compression fittings. The signal feedthroughs are 9-pin ceramic feedthroughs [12]. The signal and HV routing within the tube is accomplished with shielded kapton-insulated cable [10]. Care was taken to minimize any exposed signal conductors to the gas volume, particularly at the feedthroughs, lest ionization in the surrounding gas collect on the conductors. Such "stray ionization" which occurs away from the ceramic chambers degrades the measurements of the muon beam's spatial profile.

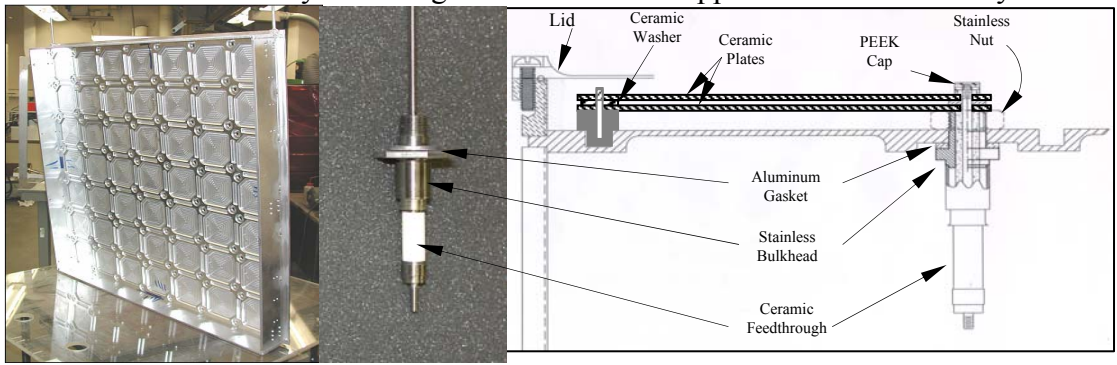

**FIGURE 5.** Construction details of the hadron monitor. (left) An Al vessel has penetrations through its rear plate for ceramic feedthroughs (middle). (right) Each ion chamber is supported on two of the feedthroughs. The front lid is a thin membrane welded to an Al flange, sealed to the vessel with an Indium wire.

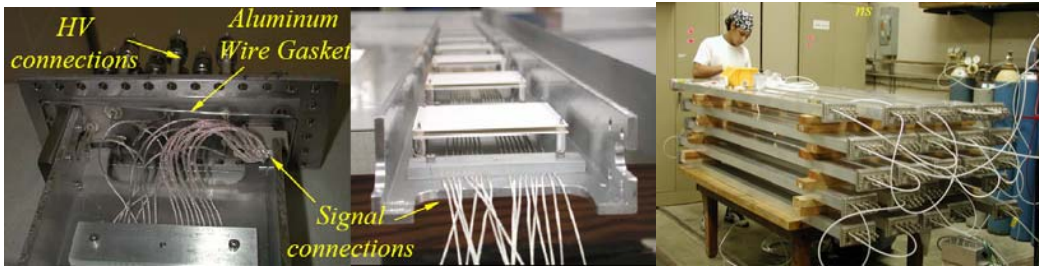

**FIGURE 6.** Construction details of the muon monitor "tubes." Each 6"x2" extruded rectangular tube contains a tray to which 9 ion chambers are mounted. Each chamber has an alpha source for calibrations. Kapton cables bring HV in and signals out. All feedthroughs are at one end of the tube.

Tests were performed of the radiation damage to detector components at the University of Texas 1 MW fission reactor. Samples of ceramic circuit boards with Ag-Pt electrodes, of kapton-insulated coaxial cable, of PEEK plastic and of aluminum compression fittings were exposed to $\sim 1.2 \times 10^{10}$ Rad. After exposure, the dielectrics could hold off >1000V with <10nA current draw. Interestingly, the kapton cable became less flexible, and the PEEK notably was of greater hardness after the test.

## BEAM TESTS

Two tests of He-filled ion chambers have been performed, both of which are described elsewhere. The first occurred at 8 GeV FNAL Booster beam dump [13], with beam intensity varied from $10^9$/spill to $4\times10^{12}$/spill. The second occurred at the Brookhaven ATF [14], where the beam was of 40 MeV electrons, intensities from $10^7$/spill to $10^9$/spill. The observed space charge effects occurred at intensities an order of magnitude larger than required for NuMI (see Table 1).

## BACKGROUND RATES

The hadron monitor and the muon detectors in Alcove 0 are expected to see, by virtue of their close proximity to the beam absorber, relatively large rates of background particles, *viz.* gamma rays and low energy neutrons. Monte Carlo estimates of the neutron rates indicate as much as a factor of 10 more neutrons than charged particles reaching those detectors. Because these neutrons are produced by the proton beam hitting the absorber, they are largely uncorrelated spatially or in intensity with the neutrino beam. We attempted to measure using radioactive sources the ionization rate from neutron-induced recoils in our chambers [15]. The results indicate a background rate from neutrons of <¼ that expected from charged particles.

## MUON MONITOR CALIBRATION

The precise measurement of muon beam profiles requires a relative chamber-to-chamber calibration within the muon monitor arrays. In fact, the uncertainties in muon beam centroid position listed in Table 1 are dominated by the chamber-to-chamber calibration uncertainty, not by the inter-chamber spacing or the number of chambers in the array. We therefore require a 1% chamber-to-chamber calibration.

The calibration procedure was to irradiate each of the detectors within the muon tubes with a source (1 Ci $^{241}$Am, 30-60 keV γ rays). The relative ionization currents induced by this source equals the relative calibration factor to be applied to each of the 243 chambers within the muon system. All 9 chambers within a tube were read out simultaneously during one tube's calibration, so that the relative sizes of the signals from the 1 μCi α sources could be studied as well. We monitored temperature and absolute gas pressure within the tube with dedicated sensors to correct for any variations over the course of testing all the tubes. As a check of these sensors, a "reference chamber" was constructed and placed in series with the tube's gas flow. This reference chamber had a 40 μCi α source mounted inside it to act as a standard signal which should change only if gas pressure, temperature, or purity level changes.

The calibration test stand is shown in Figure 7, and some of the results in Figure 8. Bias voltage curves were recorded for each chamber irradiated by the 1 Ci source, and plateau ionization currents noted. As a check of the calibration procedure's precision and repeatability, as well of that of the pressure and temperature monitoring, we measured one particular chamber approximately 100 times, as shown in Figure 8. The repeatability is 0.2pA/123.0pA≈0.2%, adequate for our 1% calibration requirement. A histogram of the plateau ionization currents of all the calibrated chambers is shown in the right plot of Figure 8. The 20 pA variation in chamber plateau currents reflects differences in the assembly of each chamber, such as the electrode spacing.

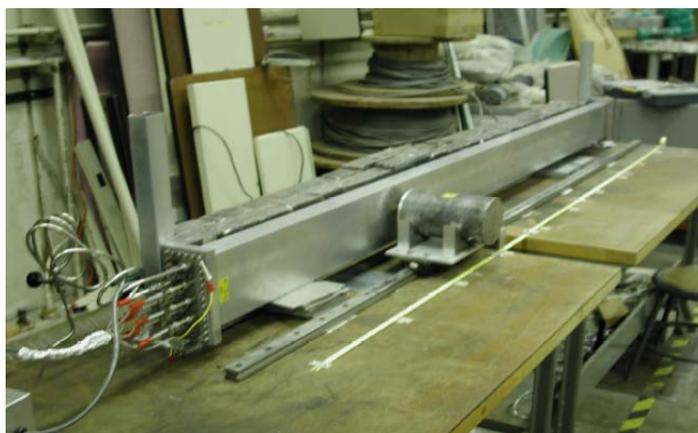

**FIGURE 7.** Muon monitor calibration test stand. The muon monitor tube is mounted on a fixed structure, in front of which is sanned a 1 Ci $^{241}$Am γ-source.

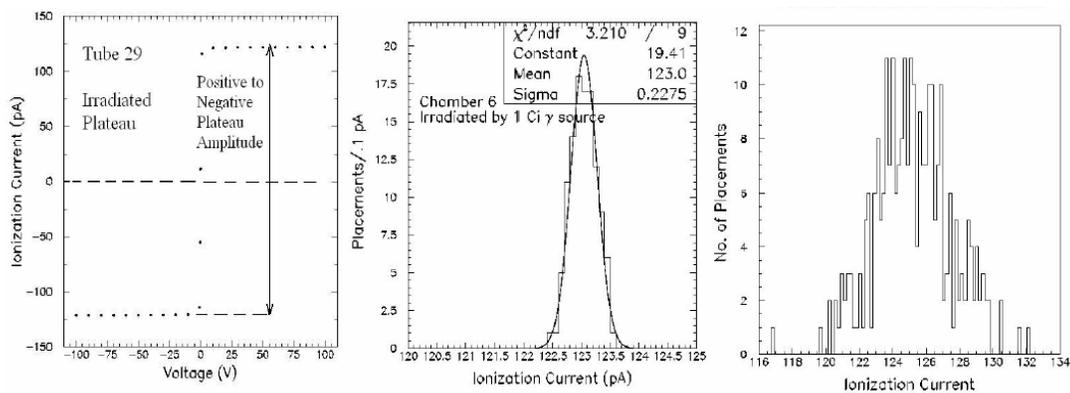

**FIGURE 8.** (left) Bias voltage curve of one chamber illuminated by the 1 Ci calibration source. (middle) Study of repeatability of 100 calibrations of the same chamber over the course of 8 hours. (right) Ionization currents of all 288 chambers calibrated on the test stand.

## ACKNOWLEDGMENTS

It is a pleasure to thank S. O'Kelley of the University of Texas Nuclear Engineering Teaching Center, the staff of the University of Texas Physics Department Mechanical Shops, K. Kriesel of the University of Wisconsin Physical Sciences Laboratory, and B. Baller, R. Ducar, D. Pushka, G. Tassotto, and K. Vaziri of Fermilab for valuable collaboration on this project.